# Observation of strain-free rolled-up CVD graphene single layers: towards unstrained heterostructures

*Ingrid D. Barcelos†\*, Luciano G. Moura‡, Rodrigo G. Lacerda†, Angelo Malachias†\**

†Departamento de Física, Universidade Federal de Minas Gerais (UFMG), Avenida Presidente Antônio Carlos 6627, 31270-901 Belo Horizonte, MG, Brazil,

‡Departamento de Física, Universidade Federal deViçosa (UFV), Avenida Peter Henry Rolfs, s/n - Campus Universitário, Viçosa - MG, 36570-000, Brazil

\*Address to corresponding authors: ingridfisica@gmail.com and angelomalachias@gmail.com

ABSTRACT

Single layer graphene foils produced by Chemical Vapor Deposition (CVD) are rolled with self-positioned layers of InGaAs/Cr forming compact multi-turn tubular structures that consist on successive graphene/metal/semiconductor heterojunctions on a radial superlattice. Using elasticity theory and Raman spectroscopy we show that it is possible to produce homogeneously curved graphene with curvature radius on the 600nm-1200nm range. Additionally, the study of tubular structures also allows the extraction of values for the elastic constants of graphene that are in excellent agreement with elastic constants found in the literature. However, our process has the advantage of leading to a well-defined and nonlocal curvature. Since our curvature radius lie in a range between the large radius studied using mechanical bending and the reduced radius induced by Atomic Force Microscopy experiments we can figure out whether bending effects can be a majoritary driving force for modifications in graphene electronic status. From the results described in this work one can assume that curvature effects solely do not modify the Raman signature of graphene and that strain phenomena observed previously can be ascribed to stretching due to the formation of local atomic bonds. This implies that the interactions of graphene with additional materials on heterostructures must be investigated in detail prior to the development of applications and devices.

The structure of graphene, a two-dimensional layer of regularly arranged carbon atoms with hexagonal structure, has been responsible for an upheaval in scientific expectations ranging from basic physics [1, 2] to applied material engineering [3, 4]. In particular, the possibilities of tuning graphene properties by doping [5], piling up a fixed amount of layers [6] and straining its crystalline structure [7] increased considerably its versatility and capabilities for future applications. Additionally, a general challenge is also imposed by the intention of integrating graphene with current electronic technologies. For such purpose, the production of heterostructures with well-defined interfaces is mandatory. A special type of self-assembled object ranging from micrometer to nanometer size such as a rolled-up tube can be used to produce the desired integration [8, 9]. These objects are obtained by strain engineering of thin flat layers in which the built-in interface properties can be directly modified to induce the formation of tubes with deterministic morphologies and positioning. In general terms a rolled up tube can be made out of two or more stacked layers grown on top of an etchant sensitive film. By inducing a well-defined strain in each interface of this system the films will curve with a characteristic radius, which minimizes elastic energy, once the etchant sensitive (sacrificial) layer is chemically removed [10]. The number of turns produced depends on the etching speed and time, which renders the technique suitable to generate single wall objects as well as artificial radial superlattices [11]. Several applications can be then envisaged for microelectronics [12], optics [13] and material junctions and heterostructures that cannot be produced by conventional deposition methods [14]. In this work we have produced rolled up tubes consisting of an InGaAs/Cr bilayer with and without graphene single layers. The knowledge of the original bilayer system allows a better understanding of the structures to which large area CVD-grown graphene has been transferred. Our technique allows the production of an array of tubular graphene heterostructures with controllable and homogeneous curvature without introducing defects or inducing strain by local chemical interactions. It is known that a monolayer structure submitted to strain can have its electronic and optical properties considerably modified [15]. First principle calculations point out to the opening of a gap in the K point of graphene by applying a uniaxial stress of the order of 0.01 (1%) [16]. In such scenario the gap can be tuned by adjusting an external stress and its consequent strain. An alternative suitable scenario is also achieved by imposing a well-known curvature to graphene, either as a method to improve its surface/volume ratio or as an effective way to produce tangential strain [17].

Past works have managed to roll either exfoliated graphene [18] or produce tubular structures by chemical synthesis using thin Cu wires as templates [19]. Our approach has the advantage of controlling the interface types due to the materials deposited for the rolling process as well as the number of piled graphene layers with reduced amount of defects, spanning over a large area. One can envisage that keeping or modifying such tubular graphene heterostructures in a controlled fashion would be desirable for sensors [20], electronic devices such as capacitors [21] and transistors [22], electromechanical actuators [23] and intelligent systems that respond to external stimulation [24]. Finally, curving graphene layers is a valuable tool to evaluate elastic properties [25], especially if a non-local curvature (extending homogeneously over a large area) can be produced.

For all tubes rolled in this work, with or without graphene, a similar layer stack was used. On the top of a GaAs (001) substrate, a 20nm AlAs layer was grown, followed by a 15nm $In_{0.2}Ga_{0.8}As$ layer, both deposited by Molecular Beam Epitaxy (MBE) at NOVA electronic Materials (Flower Mound, USA). Periodic lithographic stripes with 100 μm, separated by 50 μm, were defined by optical lithography along the [100] direction. Chromium layers with selected thickness were then thermally evaporated on top of the patterned surface. After a lift-off process the Cr film remains at the 100 μm stripes [Fig. 1(a)]. At this point the processing of tubes of InGaAs/Cr and tubes where a graphene monolayer is added differs. If one wishes to produce only InGaAs/Cr tubes, a second lithographic step consisting of shallow etching is applied to the exposed 50 μm stripes, using a solution of $H_3PO_4(85\%) : H_2O_2(31\%) : H_2O$ (1:10:500), removing the InGaAs film and exposing the AlAs sacrificial layer [26] as represented in Fig. 1(c). At the positions where Cr was evaporated the local in-plane GaAs lattice parameter is preserved since the Cr lattice parameter along the [110] direction is equivalent to approximately half of the GaAs lattice parameter in the same direction [27]. By removing the AlAs layer, with a diluted $HF(50\%) : H_2O$ (1:800) solution during 8 minutes [Fig. 1(d)], rolled-up tubes with radius ranging from 650nm to 1050nm were produced (varying according to the Cr thickness). In order to roll tubes with a graphene single layer an additional step is necessary: the graphene film has to be transferred to the top of the Chromium layer as shown in Fig. 1(b), and the definition of stripes containing graphene on top is made using lithography and Oxygen plasma [Fig. 1(c)]. The final tubular structure obtained, rolled from one etching front located at a trench is represented in Fig. 1(e).

Our tubes were defined to have 100 μm length and were found to be homogeneous after roll, as shown in Fig. 2. The tubes obtained with graphene layers have the same quality of their counterparts without graphene, exhibiting, however, a larger radius. The single-layer graphene used in this work were grown by CVD on top of Cu foils (25 μm). The procedure used for graphene formation consists on heating the Cu foil to 1000°C with 60sccm $H_2$ flow at 730mtorr, followed by a 40min annealing at this temperature. A mixture of $CH_4$(33% volume) and $H_2$(66% volume) is inserted into the furnace, and kept at 330mtorr along 2.5 hours. Finally, the sample is cooled to room temperature under $CH_4$ : $H_2$ flow and transferred to the top of the rolled-up tube layer stack using a protective polymeric poly(methylmethacrylate) (PMMA) [28]. It must be mentioned here that the graphene transfer process includes the removal of the underlying copper using an $(NH_4)_2S_2O_8$ : $H_2O$ solution (0.1 mol/l), which does not degrade the elastic properties of graphene according to ref. [29].

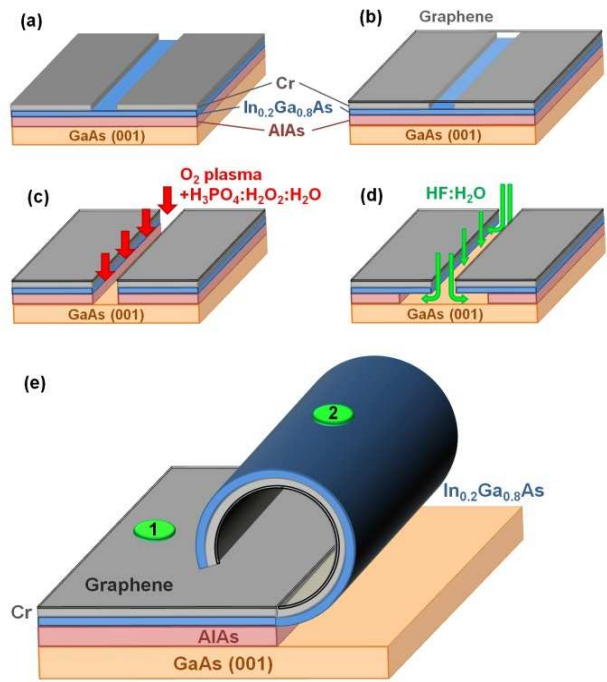

Fig. 1 – Steps for the production of rolled up tubes of InGaAs/Cr/graphene. (a) On the top of a GaAs (001) substrate an epitaxial layers of AlAs (20nm) and $In_{0.2}Ga_{0.8}As$ (15nm) are grown by MBE, followed by the deposition of a Cr thin film by thermal evaporation. Periodic Chromium stripes are defined by optical lithography. (b) CVD-grown graphene single layers can then be transferred to the top of the Chromium stripes. (c) Selective etching of graphene is carried out by Oxygen plasma (additional photolithography steps are not shown separately) keeping the graphene layers on top of the Cr stripes. The vertical corrosion removing the InGaAs film and exposing the AlAs is completed by wet etching in $H_3PO_4:H_2O_2:H_2O$ solution. (d) The sacrificial layer is chemically etched by diluted $HF:H_2O$ solution and (e) the film curls, producing tubes with well-defined radius.

In order to visualize the formed tubes and evaluate the rolling radius, as well as the incorporation of graphene, scanning electron microscopy (SEM) measurements were carried out.

For all tubes, 100 µm long homogeneous cylindrical structures were obtained, with few cracks localized at the tube openings. Figure 2(a) shows a top view of a InGaAs (15 nm) / Cr (17.5 nm) tube ensemble, showing typical tubes that roll over 30 µm distances (determining, therefore the number of turns according to the resulting tube radius). Figure 2(b) shows a single tube of the same InGaAs/Cr bilayer, but containing a single graphene layer. A detailed view of the tube opening is shown in panel (c). One notice here that the successive windings are very compact for tubes with [Fig. 2(c)] and without graphene [Fig. 2(d)]. A direct visualization of graphene rolled layers was possible in tubes where small cracks were present at the tube opening, as shown in Fig. 2(e). In all tubes containing the CVD graphene it was possible to observe smooth layers with good interfaces and absence of folded or wrinkled regions, attesting the good quality of the rolling process. Hence, the graphene layers processed by this method are subjected to a fixed curvature along a large area and protected from modifications due to ambient exposure.

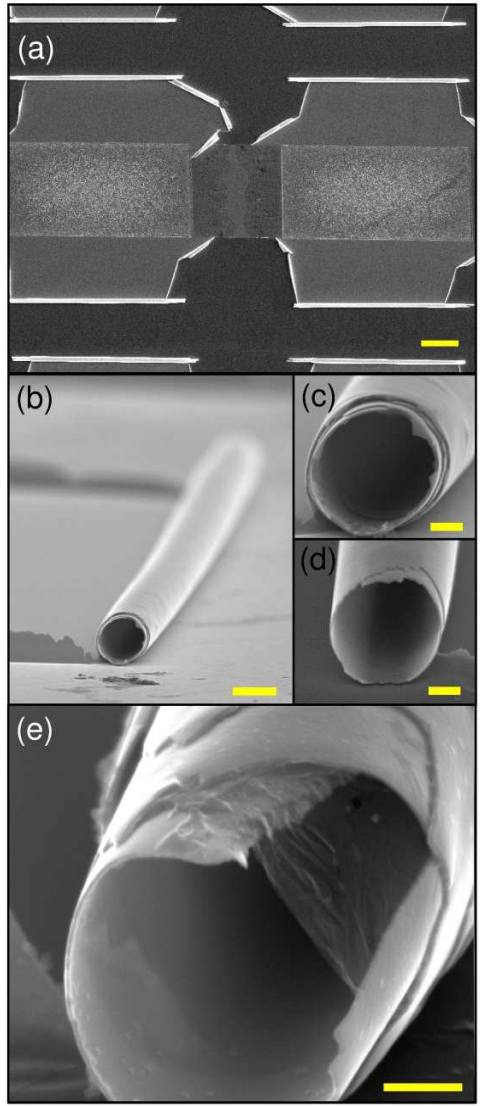

Fig. 2 – Scanning Electron Microscopy images of selected InGaAs/Cr (17.5 nm) and InGaAs/Cr (17.5 nm) /graphene rolled-up tubes. (a) Upper view of the resulting tubes on the sample surface after an optical lithography process (tubes have 100 µm length). (b) View of a single tube with graphene, and detail of the tube opening (c). (d) Tube opening for InGaAs/Cr rolled layers. (e) Magnification of a tube opening showing a graphene layer extending over a local crack. The yellow scale bars are equivalent to 20 µm in (a), 2µm in (b) and 500nm in (c, d, e).

At this point one must find out if successive windings of the tubes are physically compact, keeping the unwrinkled graphene layers inside the tubes. Such issue can be tackled by statistical analysis of the internal and external radius of the tube ensembles. For each sample (with or without graphene) 50 tubes or more were analyzed by SEM. Figure 3 summarizes the statistical data for tubes produced with a fixed Cr layer thickness of 9.2nm. Tubes produced without graphene perform between 2.5 and 3 turns and exhibit an average internal radius of 653nm while tubes with graphene perform between 2 and 2.5 turns with internal radius of 718nm. In Fig. 3(a) we show the distribution of internal radius and external radius evaluated by SEM for tubes without graphene. One can observe that the distribution of internal radius is narrower than the distribution of external radius due to a fraction of tubes that performed half-complete turns. A Gaussian fit (solid line) yields 6 nm radius dispersion, indicating a very deterministic radius for the produced structures. As layers roll, the observed external radius increases, as well as the radius distribution, which has a dispersion of about 11nm. While the average value for external radius point out to compact tubes, since the increase in this value is directly related to the number of turns, the dispersion (about 9nm from the Gaussian fit) indicates that there is a slight fluctuation of the total number of turns). For tubes rolled with graphene the statistical analysis shown in Fig. 3(b) also yields narrow distributions, but with a slight increase in the Gaussian fit widths: 14 nm for both internal and external radius. Again, the difference in average radius from the internal to the external radius is a multiple of the total layer thickness and allows us to infer that the successive windings are compact. Such analysis was repeated for each sample studied here, giving rise to the data points of Fig. 4, discussed in the following paragraph. This figure also points out that large arrays of tubes with controllable radius can be produced using our method.

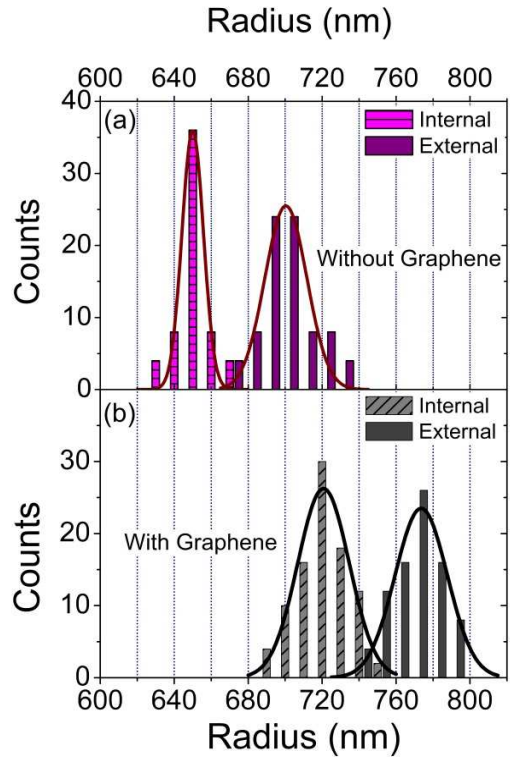

Fig. 3 – Histograms of internal and external radius measured by SEM on InGaAs/Cr and InGaAs/Cr/graphene tubes produced from layers with a 9.2nm Cr thickness. The results for tubes without graphene are shown in (a) whereas the results for tubes with graphene are depicted in (b). Gaussian fits are drawn as a guide to the eyes, allowing an estimation of the radius dispersion in each case.

In order to evaluate the graphene elastic constants in our system it is necessary to take into account that the driving force for rolling in our structures comes from the strain in the InGaAs/Cr interface. The polycrystalline Cr keeps the GaAs in-plane lattice parameter of the substrate, making the bilayer system behave similarly to an InGaAs/GaAs bilayer [30, 31], but with different elastic constants at the upper (Cr) layer. It is mandatory, therefore to extract the Cr properties prior to extending the analysis to tubes containing graphene.

Using an analytical solution depicted in ref. [31] one can directly retrieve the strain status and elastic properties of the Cr thin films from a reference sample series with different Cr thickness. In our case, the rolling radius is a function of the strain - kept as a value of 0.0137, close to the GaAs/In$_{0.2}$Ga$_{0.8}$As - and elastic constants of Cr such as Poisson's ratio (ν) and Young Modulus (E). From the dotted line fit to the open symbols (data) in Fig. 4 we could extract ν = 0.28(1) and E = 109(5) GPa for the Cr thermally evaporated polycrystalline films. Using such values the treatment of the rolled layers with graphene is carried out assuming an interface Cr/graphene free of strain (no crystalline registry), a graphene thickness of 0.4nm and the Cr parameters were extracted from the first sample series. From such approach to the sample series with graphene single layers one obtains the solid line fit to the solid dots (data) shown in Fig. 4. A Young modulus of 800(20) GPa is retrieved, close to the value of exfoliated graphene, as well as continuous CVD graphene films with small grains and isolated single-crystal graphene [29]. An effective in-plane Poisson's ratio of 0.20(0.03) was found, near values obtained by other groups for free-standing single layers [32, 17]. One must notice that this first quantitative result lies on the analysis of a reliable constant-curvature procedure, which is more homogeneous and has the advantage of allowing a directly evaluation by continuous elasticity theory. Experimentally, the Young modulus of CVD-graphene has been measured by AFM [33, 25], where the surface curvature and local effects coming from inhomogeneities in composition and lattice (defects) of the graphene foil, as well as inaccuracies considering the exact contact area due to morphological characteristics of the AFM tip, produce deviations from results obtained in different systems, yielding a broad range of values. For single layers typical AFM experiments as well as theoretical works point out to values for E spanning from 500GPa up to 1120GPa, reaching 6.9TPa for multilayers [34, 33]. Hence, we show here a large-area measurement in opposition to extremely localized results obtained using a local probe.

In the precedent discussion the absence of strain on the rolled graphene layers were assumed in order to provide a suitable fit to the observed tube radius (Fig. 4). Although this is an indirect evidence of reduced strain one can directly probe such fact using Raman spectroscopy. Raman measurements were performed using a Renishaw InViamicro-Raman system. The excitation wavelength was fixed to 514.5nm and the laser power at the sample position set to 0.5 or 0.1mW, resulting in a laser spot size around 1 µm diameter. The scattered light from the sample was collected in back scattering geometry using a 50X (N. A.= 0.75) objective lens.

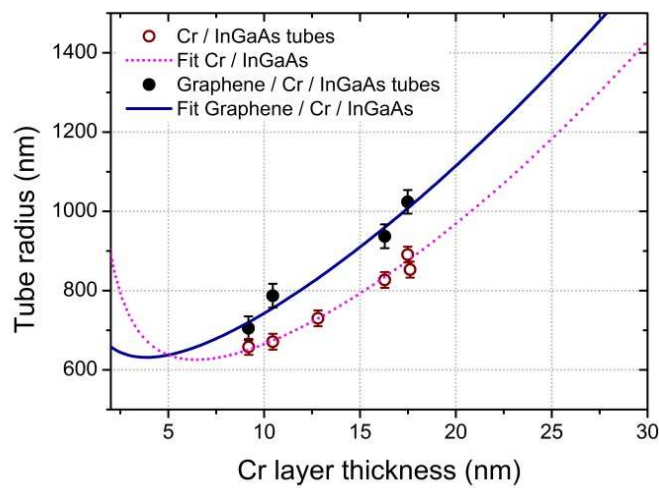

Fig. 4 – Analysis of the tube radius for the InGaAs/Cr and InGaAs/Cr/graphene rolled-up tubes as a function of the deposited Cr thickness. The symbols are measured tube radius: open dots for InGaAs/Cr and solid dots for tubes with graphene. Lines are fits using a continuous elasticity multilayer model [31] for each case.

In our layered structure graphene peaks appear far away from Raman peaks from GaAs (around 291 cm$^{-1}$), AlAs (399 cm$^{-1}$) and In$_{0.2}$Ga$_{0.8}$As (285 cm$^{-1}$) [35, 36]. Additionally, no peak was observed for Chromium oxide, indicating that most of the metal layer remains chemically unchanged. The results for the flat layers are compatible with unstrained CVD graphene with a very reduced amount of defects (the D band is very weak), as observed from the peak positions and widths. The rolling process does not create defects since the band widths remain unchanged and no additional intensity is observed at the D band for positions inside the tubes. The full Raman spectra for flat and rolled graphene layers are shown in Fig. 5(a).

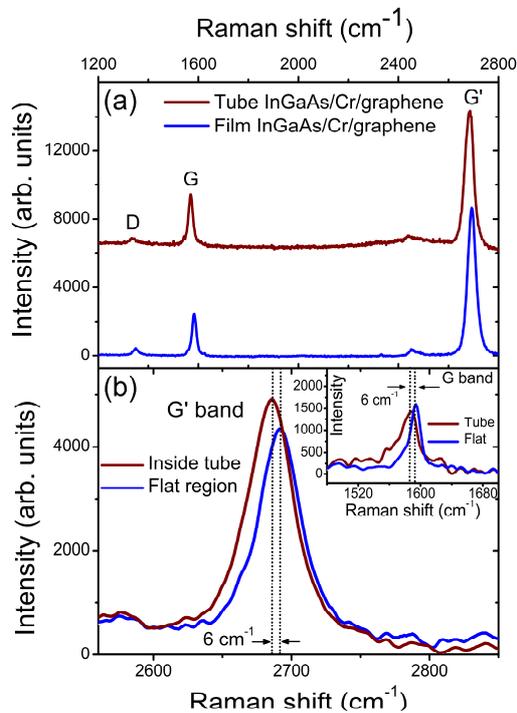

Fig. 5 – (a) Raman spectra showing all graphene-related peaks for flat and rolled layers. (b) Detailed zoom of the Raman spectra of (a) showing the G (inset) and G' bands of the graphene at the flat surface [point labeled 1 in Fig. 1(e)] and on top of the tube [point labeled 2 in Fig. 1(e)], respectively, for the samples used in this work. The spectra show the maximum band shift observed for the R = 787nm tube.

Figure 5(b) shows the maximum shift observed for the G and G' bands of graphene comparing their positions on the flat (unrolled) [represented as the position labeled 1 in Fig. 1(e)] regions as well as on top of the tubes [Fig. 1(e), position labeled as 2] with R = 787nm. In contrast with ref. [17] where curvature radius of a few centimeters lead to a large band shift, for the small curvature radius of our experiments only reduced redshifts below 6 cm$^{-1}$ for the G and G' band were observed. Using the model of ref. 17 the strain deduced from such deviations according to the relation explored in [17] can range from 0.002 to 0.004, depending on the Grüneisen parameter γ used (parameter that describes frequency shifts originated by hydrostatic strain). Nevertheless, the shifts observed are extremely small and, independently from the curvature model assumed to analyze the Raman shifts the resulting strain retrieved will always be very reduced.

A clear demonstration of the small shifts observed in this work is found in Fig. 6, where we take the fitting and selected data from ref. [17] and reproduce it in the left panel, while our data is exposed at the right panel. The data shown in the left panel is independent of a strain model since curvature radius and band shifts can be directly measured. This is also the case for our data on the right panel, where average values over more than 10 tubes for each curvature radius were plotted as solid symbols, and the maximum shift for each tube radius is shown as open symbols. One can notice that although a clear trend of the data points is not observed, the measured shifts are extremely reduced for the bend radius retrieved in rolled-up tube systems, indicating that the strain induced on graphene layers cannot be a result of an externally imposed curvature radius.

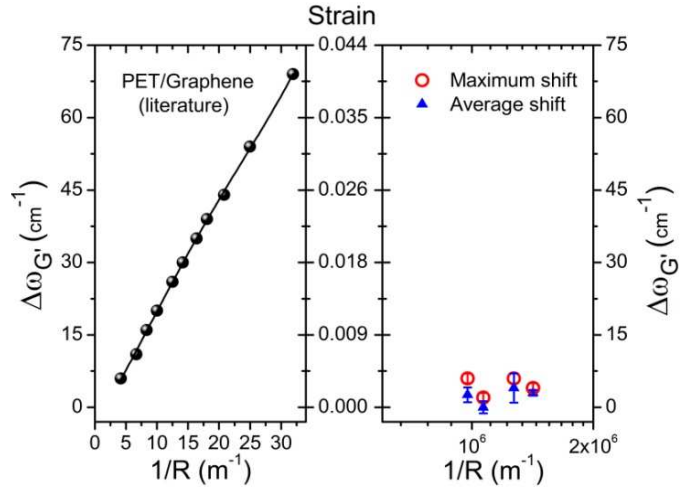

Fig. 6 – Left panel: shift of the G' band with respect to unstrained graphene as a function of the inverse of the curvature radius for mechanically bent exfoliated graphene monolayers. The fit and selected data were extracted from figure 3 from ref. [17]. Right panel: shift of the G' band for rolled-up graphene CVD single layers studied in this work. The maximum and average shifts are directly displayed by the open and solid symbols, respectively.

The results of figures 5 and 6 suggest that the strain-induced shifts in Raman bands G and G' for graphene single layers have chemical origin. In other words, the electronic structure of graphene may change upon curvature – and therefore show a signature in Raman measurements – depending whether the graphene has any chemical affinity with the materials that are in contact with it. Although this conclusion sounds trivial at a first glance, it inherently implies that the interactions of any material that has to be integrated with graphene must be investigated in detail prior to the development of systems from which a given response is expected (e.g., devices or applications). In our case, the monolayer of graphene is sandwiched between Chromium and InGaAs (in contact with it after rolling), and our results indicate that very weak interactions take place among these layers. The fact that reduced amounts of strain are observed points out that one can roll up graphene with small curvature radius excluding strong structural deformations. An important indication of this weak interaction among layers is the value retrieved for the Poisson's ratio, which was found to be similar to free-hanging graphene [32, 17]. The results explored in ref. [17] make use of ν = 0.33, compatible with the case of bonded layers.

In summary, our technique allows the production of controlled and homogeneous curvature in graphene layers without introducing defects or inducing strain by local chemical interactions. The tubes shown here represent a realization of a graphene/metal/semiconductor radial superlattice, creating heterojunctions that cannot be easily piled up by other methods. This opens up the possibility of spin injection devices using oriented unstrained graphene [37] to produce a conduction channel which is well-defined along the rolled-up tube axis. The rolled-up system also optimizes the volume-area ratio of graphene layers without modifying properties due to direct stacking or folding, which is also of interest for devices such as sensors and capacitors. Futhermore, we were also able to retrieve values for the elastic constants of graphene through a nonlocal (more homogeneous) bending method yielding reduced error bars with respect to other methods. These values are in excellent agreement with graphene elastic constants found in the literature. We believe that the integration of graphene with other materials has to be thoroughly investigated, on both experimental and theoretical aspects, in order to reveal whether local interactions impact on electronic properties. Finally, we have show that curvature effects solely do not modify the Raman signature of graphene down to radius of about 600nm.

The authors acknowledge CAPES, FAPEMIG and CNPq for financial support, as well as INCT-Carbon. The use of the Electron Microscopy Center at UFMG is also greatly acknowledged.